\documentclass[aps,prb,twocolumn,groupedaddress,showpacs,amsmath,amssymb]{revtex4}

\usepackage{amsfonts}
\usepackage{amssymb,amsmath}
\usepackage{graphicx}
\usepackage{dcolumn}
\usepackage{bm}
\usepackage[dvips]{color}

\newcommand{\be}{\begin{equation}}
\newcommand{\ee}{\end{equation}}
\newcommand{\bea}{\begin{eqnarray}}
\newcommand{\eea}{\end{eqnarray}}
\newcommand{\la}{\langle}
\newcommand{\ra}{\rangle}
\newcommand{\ld}{\left(}
\newcommand{\rd}{\right)}

\newcommand{\lbr}{\left[}
\newcommand{\rbr}{\right]}
\newcommand{\non}{\nonumber \\ }
\newcommand{\ve}{\varepsilon}

\begin{document}

\title{Exact results for spatial decay of correlations
in low-dimensional insulators II}

\author{Janusz J\c{e}drzejewski$^{1}$}
\email[Electronic address: ]{jjed@ift.uni.wroc.pl}
\author{Taras Krokhmalskii$^{2,1}$}
\email[Electronic address: ]{krokhm@icmp.lviv.ua}

\affiliation{        $^{1}$Institute of Theoretical Physics,
                    University of Wroc\l aw, 
        pl. Maksa Borna 9, 50-204 Wroc\l aw, Poland\\
        $^{2}$Institute for Condensed Matter Physics, 
        1 Svientsitskii Str., L'viv-11, 79011, Ukraine}


\date{\today}



\begin{abstract}
We study decay rates of one-body reduced density matrices in
insulators, described by a tight-binding model, where not only an
external potential but also hoppings are spatially modulated. We
determine analytically the power in the power law and the
correlation length in $D=1$ case and in several lattice directions
in $D=2$ case. Unlike the uniform hopping case, in $D=1$ system
and in some directions of $D=2$ system the correlation length is
not determined uniquely by the gap. Moreover, a crossover from
$D=2$-decay rates to $D=1$ ones is investigated.
\end{abstract}

\maketitle

\newpage

The localization properties of one-body reduced density matrices
(DM) are of vital importance for rapidly developing computational
techniques, used for calculating various characteristics of
crystalline solids \cite{payne,goed1,kohn1}.
But the interest in large-distance decay
rates of DM dates from 1959, when Kohn \cite{kohn2} determined an
exponential factor in the large-distance asymptotics of DM in
one-dimensional ($D=1$) crystals. Recent works concerning this problem
include Refs.\cite{baer,goed2,beigi,he,zhang,taraskin1,taraskin2,jk}.
In particular, another major step, after Kohn's result, was made
by He and Vanderbilt who discovered a power-law factor in the
dacay rate of $D=1$ crystals.
Extensions to systems in dimensions $D>1$ have been obtained by
Des Cloizeaux \cite{cloizeaux},
Taraskin et al \cite{taraskin1,taraskin2}, and in our recent paper
\cite{jk}, where we have shown to what extent
the large-distance asymptotics of DM is sensitive to lattice directions
in $D=2$ crystals.

In all the papers quoted above, the pattern of hoppings is spatially uniform.
Here, we consider a system of spinless fermions on a lattice whose hoppings
are spatially modulated, subjected to a modulated external potential.
The Hamiltonian of the considered system reads:
\be
\label{H}
H = \sum_{\bm i} U_{\bm i}a^+_{\bm i} a_{\bm i} +
\sum_{\la \bm i,\,\bm j \ra} t_{\bm i \bm j}\ld a^+_{\bm i}a_{\bm
j}+h.c. \rd .
\ee
In the above expression, $\bm i,\bm j$ represent the lattice sites
of a $D$-dimensional simple cubic lattice, $\la \bm i,\bm j \ra$
stand for the pairs of nearest neighbors on this lattice, and in the
sum over nearest neighbors each pair is counted once. The operators
$a^{+}_{\bm i}$ $(a_{\bm i})$ create (annihilate) a
spinless fermion in a single-particle orbital $|\bm i \ra$,
belonging to an orthonormal basis. As ${\bm j}$ runs through the
underlying lattice, the value of the external potential at site $\bm
j$ is modulated with the amplitude $u$,
\be
\label{U}
U_{\bm j} = U + (-1)^{\sigma_{\bm j}} u,
\ee
where for a site ${\bm j}=(j_1, ...,j_D)$, $\sigma_{\bm j}=
\sum_{l=1}^{D} |j_l|$ is the (noneuclidean) distance of this site
from the origin. Since a $D$-dimensional simple cubic lattice
consists of two interpenetrating sublattices: the even one -- with
$\sigma_{\bm j}$ even, and the odd one -- with $\sigma_{\bm j}$ odd,
the external potential is uniform at each sublattice. Its value
changes by $\pm 2u$ on passing from one sublattice to the other one,
and the average over the whole lattice is $U$.
Similarly, the symmetric hopping matrix $t_{\bm i \bm j}$ is
modulated:
\be
\label{t}
t_{\bm i \bm i \pm \bm e_l} =
t \left[ 1 \pm (-1)^{\sigma_{\bm i}} \tau \right],
\ee
with the mean $t$ and the amplitude $t\tau$,
where ${\bm e_{l}}$, $l=1, ...,D$, is the unit vector
with only one nonzero component -- the $l$-th one.
Without any loss of generality, we can suppose that $u \geq 0$,
$t \geq 0$, and $1 \geq \tau \geq 0$.

The considered model can describe electronic properties of crystal surfaces,
possibly containing adsorbed atoms. The pattern of hoppings (\ref{t})
may arise when atoms form a distorted square lattice,
that is one of the sublattices is shifted with respect to the other.

Under the periodic boundary conditions, the Hamiltonian (\ref{H}) is
block-diagonalized by the plane wave orbitals, $|\bm k \ra$, with the
wave vector in the first Brillouin zone of the lattice.
Specifically, shifting zero of the energy scale to $U$ and
expressing all the energies involved in the units of the average
transfer integral $t$, we obtain the upper, $\Lambda^+_{\bm k}$,
and the lower, $\Lambda^-_{\bm k}$, bands of eigenenergies:
\be
\label{cbdisp}
\Lambda^{\pm}_{\bm k}=\pm \sqrt{u^2+ \ve_{\bm k}^2 +
\zeta_{\bm k}^2} \equiv \pm
\Delta_{\bm k }.
\ee
In Eq.(\ref{cbdisp}), the wave vector ${\bm k}$  is restricted to the
first Brillouin zone of one of the sublattices
(so that there is as many eigenvalues as sites in the lattice),
$\ve_{\bm k} = 2\sum_{l=1}^{l=D}\cos k_l$, $\zeta_{\bm k}=2\tau
\sum_{l=1}^{l=D}\sin k_l$. The two bands are mutually symmetric
about zero and are separated by a gap $g_D$. If the system given by
(\ref{H}) is half-filled and $g_D > 0$, then it is an insulator
in the sense of the standard band theory. The normalized
eigenvectors corresponding to the eigenvalues $\Lambda^{-}_{\bm k}$
are linear combinations of the vectors $|\bm k\ra$ and $|\bm k + \bm
\pi \ra$, where $\bm \pi$ stands for the vector with all the components
equal to $\pi$. By means of these eigenvectors, one can calculate
the zero-temperature non-diagonal elements of DM for the half-filled system,
which in the thermodynamic limit read
\bea
\label{DM}
\la a^+_{\bm i}\!a_{\bm i+ \bm r}\! \ra \! \!\!\! &=&
\!\!\! \!-\! \frac{1}{2}\!\sum_{l=1}^{D}
\! \left[\! {\cal{S}}_l {\cal{R}}_D(\bm r)\! -\!
\tau (-1)^{\sigma_{\bm i}}{\cal{D}}_l {\cal{R}}_D(\bm r)\! \right]\!, \non
&&\qquad \qquad \qquad \qquad  \mbox {if} \  \sigma_{\bm r}\!=\!2m\!+\!1,
\ \ \mbox {and} \non
\la a^+_{\bm i}\!a_{\bm i + \bm r}\! \ra \!\! \!\!&=&
\!\!\! -\frac{u}{2} (-1)^{\sigma_{\bm i}} {\cal{R}}_D(\bm r),
 \ \  \mbox {if} \  \sigma_{\bm r}=2m ,
\eea
where
\vspace{-0.5cm}
\be
\label{Rcb}
{\cal{R}}_D(\bm r)=2(2\pi)^{-D}
\int_{B.Z.} d\bm k \exp\ld i\bm k \bm r\rd  \Delta^{-1}_{\bm k},
\ee
and ${\cal{S}}_1 {\cal{R}}(\bm r) \equiv {\cal{R}}(r_1 +
1,\ldots,r_D) + {\cal{R}}(r_1 - 1,\ldots,r_D)$, ${\cal{D}}_1
{\cal{R}}(\bm r) \equiv {\cal{R}}(r_1 + 1,\ldots,r_D) -
{\cal{R}}(r_1 - 1,\ldots,r_D)$, and so on for $l=2,\ldots, D $. In
Eq. (\ref{Rcb}), the D-dimensional integral is taken over the first
Brillouin zone of a sublattice and $\bm r \neq 0$. At odd sites
($\sigma_{\bm r}=2m+1$), the function ${\cal{R}}_D$ vanishes.

At the level of the Hamiltonian (\ref{H}), it is clear that for
$\tau=0$ we are back in the the uniform hopping case of Ref.\cite{jk}.
Note however,
that here $\Delta_{\bm k}$ is twice as large as in Ref.\cite{jk},
and consequently an extra factor 2 appears in the definition of
${\cal{R}}_D(\bm r)$.
On the other hand, if $\tau=1$, the system looses its
connectivity with respect to the hopping. Specifically, for $D=1$,
we obtain a system of independent dimers whose correlation functions vanish
at distances greater than one,
while for $D=2$ -- a system of independent
quasi-one-dimensional zig-zag chains.

In what follows, we shall investigate the large-distance decay rate
of the function ${\cal{R}}_D(\bm r)$ in $D=1,2$ systems.


It is instructive to analyze first the one-dimensional case. The
gap in the spectrum is $g_1=4\delta_{1}$, with $\delta_{1}^{2}=
\beta_1^2 + \tau^{2}$, and $\beta_1=u/2$. Apparently, the role
of the rescaled modulation amplitudes, $u$ and $\tau$, in creating the gap is
symmetric. Moreover, $g_1 >0$ unless $u=\tau=0$.
For later purposes, it is convenient to introduce, besides the rescaled gap $\delta_1$,
one more parameter describing the spectrum: for the lower band this
parameter amounts to the ratio, $\kappa_1$, of the upper edge of the
band to the lower edge, i.e. $\kappa_1^2=(1-\tau^2)/(1+\beta_1^2)$.
The pair $(\delta_1,\kappa_1 )$
can be thought of as new coordinates, equivalent to the coordinates $(\beta_1,\tau )$.
The specific combination
of $\beta_1$ and $\tau$, denoted $\kappa_1$, appears naturally in an expression
for the correlation function (\ref{DM}). Provided that $\kappa_1 \neq 0$,
the function ${\cal R}_1(\bm r)$ at even points, $\bm r=2m$, can be written as
\bea
\label{R1}
{\cal R}_1(2m)=
\frac{1}{2\pi}\frac{1}{\sqrt{1+\beta_1^2}}
\int_{-\frac{\pi}{2}}^{\frac{\pi}{2}}d k \frac{\cos(2mk)}{\sqrt{1-\kappa_1^2\sin^2 k}} \non
=\frac{(-1)^m}{\pi}\frac{1}{\sqrt{1-\tau^2}}Q_{m-1/2}\ld \frac{2}{\kappa_1^2}-1   \rd ,
\eea
where $\Gamma$ stands for the Euler $\Gamma$-function,
$F$ -- for the Gauss hypergeometric function, and $Q_{\nu}$ -- for the Legendre function
of second kind \cite{be}.
Eq.(\ref{R1}) implies the following large-$m$ asymptotics:
\bea
\label{R1asympt}
{\cal R}_1(2m) \approx \frac{(-1)^m}{\sqrt{2\pi}}\frac{(1-\kappa_1^2)^{1/4}}{\delta_1}
\frac{\exp(-2m/\xi)}{\sqrt{2m}},
\label{R1asympt}
\eea
where the inverse correlation length $\xi^{-1}$ is given by
\bea
\label{xi}
\xi^{-1}\equiv \ln  \frac{1+\sqrt{1-\kappa_1^2}}{\kappa_1}.
\eea
We note that the correlation function
$\la a^+_{\bm i}\!a_{\bm i+ \bm r}\ra$  is continuous at $\tau=0$
$\tau=1$, and $\tau=\beta_1=0$.
Out of the three above cases, the asymptotics (\ref{R1asympt}) is continuous
only at $\tau=0$, where we recover the asymptotics obtained in Ref.\cite{jk}.

The new variables $(\delta_1,\kappa_1 )$ reveal clearly
the nature of the relation between $\xi$, $\kappa_1$ and $\delta_1$.
It follows from (\ref{xi}) that $\kappa_1$ determines uniquely
$\xi$ and vice versa.
But the relation between  $\xi$ (or $\kappa_1$) and $\delta_1$ is a multi-valued one.
Specifically, to given $\delta_1$ there corresponds
an interval $[\kappa_1^{min}(\delta_1),\kappa_1^{max}(\delta_1)]$ of $\kappa_1$ values,
where
$\kappa_1^{min}(\delta_1)=\sqrt{1-\delta_1^2}$ if $\delta_1 \leq 1$, and
$\kappa_1^{min}(\delta_1)=0$ if $\delta_1 > 1$, and
$\kappa_1^{max}(\delta_1)=1/\sqrt{1+\delta_1^2}$.
Consequently,
the inverse correlation length $\xi^{-1}$ can assume any value from the interval
$[\xi^{-1}(\kappa_1^{max}(\delta_1)),\xi^{-1}(\kappa_1^{min}(\delta_1))]$.
As $\delta_1$ tends to zero, the edges of the interval tend to zero as well
but its width shrinks even faster, what results
in a fairly good localization of $\xi^{-1}$, for small $\delta_1$.
Specifically, expanding the bounds at $\delta_1=0$,
\bea
\label{ximin}
\xi^{-1}(\kappa_1^{max}(\delta_1)) \approx \delta_1-\frac{1}{6}\delta_1^3+
\frac{3}{40}\delta_1^5+\ldots ,
\eea
\bea
\label{ximax}
\xi^{-1}(\kappa_1^{min}(\delta_1))\approx \delta_1 +\frac{\delta_1^3}{3}+
\frac{\delta_1^5}{5} + \ldots,
\eea
we see that to the first order in $\delta_1$ the bounds coincide and the
indeterminacy-interval width is $\delta_1^3/2$, in the leading order.

For $\delta_1 > 1$, $\xi^{-1}$ can assume any value not less than
$\xi^{-1}(\kappa_1^{max}(\delta_1))$, and this lower bound blows up as
$\delta_1 \to \infty$.

The described above indeterminacy of $\xi^{-1}$ vanishes if one of
the parameters, $\beta_1$ or $\tau$,
is kept constant, since such conditions turn the
relations between $\kappa_1$ and $\delta_1$ into single valued functions.
Of particular interest is the behavior of $\xi^{-1}$ in two such situations,
motivated physically. Firstly, a weak modulated external potential ($\beta_1$
varies in a vicinity of zero) is imposed on a connected system with modulated
hopping ($0 < \tau < 1$, fixed). Secondly, a weak modulated hopping is imposed
($\tau$ varies in a vicinity of zero) on a system in a modulated potential
($0 < \beta_1$, fixed). In the both cases, $\delta_1$ is separated from zero.
Therefore,
to describe the behavior of $\xi^{-1}$ it is more convenient to consider the
expansions in $\beta_1$ or $\tau$, respectively.
Specifically, in the first situation
\bea
\label{xibeta}
\xi^{-1}\approx \ln \sqrt{\frac{1+\tau}{1-\tau}}
 +\frac{1}{\tau}\frac{\beta_1^2}{2}
 -\frac{1+\tau^2}{\tau^3}\frac{\beta_1^4}{8} + \ldots,
\eea
while in the second one
\bea
\label{xitau}
\xi^{-1}\approx \ln(\beta_1+\sqrt{1+\beta_1^2})
+\frac{\sqrt{1+\beta_1^2}}{\beta_1}\frac{\tau^2}{2} + \ldots \ .
\eea
Clearly, in the above expansions it is not possible to go to zero
with the fixed parameters.
To obtain the corresponding expansions about zero, we note that
the function $\kappa_1^{max}(\delta_1)= \kappa_1(\beta_1,\tau)|_{\beta_1=\delta_1,\tau=0}$.
Therefore, for $\tau=0$ the expansion in $\beta_1$ coincides with the expansion (\ref{ximin})
of the lower bound for $\xi^{-1}$. Similarly,
$\kappa_1^{min}(\delta_1)= \kappa_1(\beta_1,\tau)|_{\beta_1=0,\tau=\delta_1}$,
hence for $\beta_1=0$
the expansion in $\tau$ coincides with the expansion (\ref{ximax})
of the upper bound for $\xi^{-1}$.

Yet another interesting regime is, when a connected system ($\tau \neq 1$) is subjected to
a strongly modulated external potential. Then, the large $\beta_1$ asymptotics of $\xi^{-1}$
reads:
\bea
\label{xiasympt}
\xi^{-1}\approx \ln\frac{2}{\sqrt{1-\tau^2}} +\ln\beta_1+\frac{1+\tau^2}{4\beta_1^2}
+ \ldots \ .
\eea


The two-dimensional system differs from the one-dimensional one in
many respects, due to different geometry and connectivity of the
underlying lattice. When hopping is modulated, an important
difference appears already in the spectrum: the gap $g_2=8\beta_2$,
$\beta_2=u/4$, is independent of the amplitude $\tau$, and vanishes
unless the external potential is modulated.

The function ${\cal R}_2({\bm r})$ can be cast into the form
\bea
\label{R2}
{\cal R}_2({\bm r})\!=\! \frac{1}{\pi^2}\!\!
\int_0^{\frac{\pi}{2}} \!\!\!\! \int_0^{\frac{\pi}{2}}  \!\!\!\!
\frac{ dx dy\cos(2mx)\cos(2ny)} {\sqrt{\beta_2^2\! +\!
\lbr \tau^2\!+\!(1\!-\!\tau^2)\cos^2y \rbr \cos^2x}},
\label{R}
\eea
where we introduced the integers $m$ and $n$ such that $2m=r_1 - r_2$,
$2n= r_1 + r_2$.
For $\tau =0$, ${\cal R}_2({\bm r})$ coincides with that of Ref.\cite{jk}.
On the other hand, in the case of
quasi-one-dimensional zig-zag chains (i.e. $\tau =1$), ${\cal
R}_2({\bm r})$ is nonzero only along direction $n=0$, and in
this direction it amounts to ${\cal R}_1({\bm r})$ but with twice as
large hopping as in the uniform one-dimensional case.

By expanding the square root in the denominator,
the double integral of (\ref{R2}) can be related to
a single integral involving a hypergeometric function $_3F_2$ \cite{be}:
\bea
{\cal R}_2({\bm r})=
\frac{(-1)^m}{2\pi}\frac{(1-\tau^2)^{-n}}{(2\beta_2)^{2m+1}}
\frac{\Gamma(m+1/2)}{\Gamma(n\!+\!1/2)\Gamma(m\!-\!n\!+\!1)}\non
\times \int_{\tau^2}^1
 dvv^{m-n}(1-v)^{n-1/2}(v-\tau^2)^{n-1/2} \non \times_3 F_2 \!\ld \!
m\!+\!\frac{1}{2},m\!+\!\frac{1}{2},m\!+\!1;2m\!+\!1,m\!-\!n+\!1;\!-\frac{v}{\beta_2^2}\!
\rd ,
\label{Rm}
\eea
for $m\geq n$, and
\bea
{\cal R}_2({\bm r})=\frac{(-1)^n}{2\pi}\frac{(1-\tau^2)^{-n}}{(2\beta_2)^{2n+1}}
\frac{\Gamma(2n+1)}{\Gamma(n\!+\!m\!+\!1)\Gamma(n\!-\!m\!+\!1)}\non \times
\int_{\tau^2}^1 dv(1-v)^{n-1/2}(v-\tau^2)^{n-1/2} \non \times_3
F_2 \!\ld \! n\!+\!\frac{1}{2},n\!+\!\frac{1}{2},n\!+\!1;n+m\!+\!1,n\!-\!m+\!1;\!-\frac{v}
{\beta_2^2}\! \rd\!,
\label{Rn}
\eea
for $n\geq m$.

Using representations (\ref{Rm}), (\ref{Rn}),
we have derived large-distance asymptotic behavior of
${\cal R}_2({\bm r})$ in diagonal directions of a connected system
($\tau \neq 1$).
Specifically, in  direction $n=0$ it follows from (\ref{Rm}) that
\bea
{\cal R}_2(m,-m) \approx
\frac{(-1)^m}{2\pi\beta_2}\frac{1}{\sqrt{1-\tau^2}}\frac{\exp(-2m/\xi)}{2m},
\label{R2mm}
\eea
\bea
\xi^{-1}=\ln(\sqrt{1+\beta_2^2}+\beta_2).
\label{ximm}
\eea
Apparently, in direction $n=0$ the gap determines $\xi$.
The above asymptotics differs only by the factor $(1-\tau^2)^{-1/2}$ from that
in the uniform hopping case \cite{jk}, hence it is continuous at
$\tau=0$.
More interesting is a neighborhood of $\tau=1$, where we would like to observe
how our $D=2$ system approaches the corresponding $D=1$ system.
In particular, an interesting question is how the power law $r^{\gamma}$,
with $r=2m$ and $\gamma=1$ in (\ref{R2mm}), transforms into $\sqrt{2m}$ in
(\ref{R1asympt}), as $\tau \to 1$? Our numerical calculations suggest that
$\gamma$ behaves discontinuously: it grows to some value above 1, and then
jumps to $1/2$ at $\tau=1$.

\begin{figure}[t]
\includegraphics[clip=on,width=8.cm]{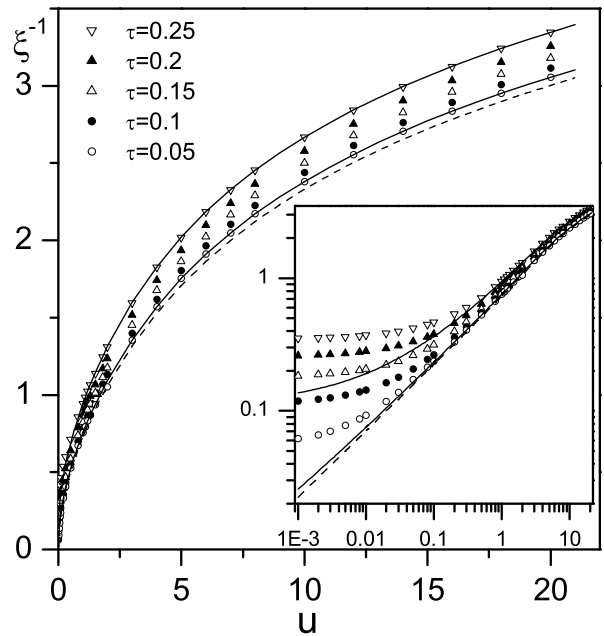}
\caption[]{The inverse correlation length of D=2 system in direction $n=m$ against $u$,
for different values of $\tau$. Symbols represent the curves obtained from fitting
numerically ${\cal R}(2m,0)$. The continuous lines are derived from
approximate formula (\ref{xiappr}),
the upper one for $\tau=0.25$, the lower one for $\tau=0.05$.
The dashed curve is derived from an exact formula for $\tau=0$.
In the inset we display the corresponding log-log plots.
}
\label{fig}
\end{figure}

Then, in direction $m=0$ formula (\ref{Rn}) gives
\bea
{\cal R}_2(n,n) \approx
\frac{(-1)^n}{2\pi\beta_2}\frac{\exp(-2n/\xi)}{2n},
\label{R2nn}
\eea
where $\xi^{-1}$ is given by (\ref{xi}) with $\kappa_1$ replaced by $\kappa_2$
and $\kappa_2$ obtains from $\kappa_1$ by changing $\beta_1$ into $\beta_2$.
Similarly, we introduce $\delta_{2}$, $\delta_{2}^{2}= \beta_2^2 + \tau^{2}$.
While $\delta_2$ and $\kappa_2$
lack the physical interpretation of $\delta_1$ and $\kappa_1$,
the formal relation between them is fairly analogous to that between the latter
parameters.
Consequently, in direction $m=0$ the gap does not determine
$\xi^{-1}$.

In directions other than diagonal ones, we have not been able
to establish the asymptotic behavior of the correlation function.
But with the help of numerical calculations, we have obtained an
approximate asymptotics in the axial direction $m=n$.
This asymptotics can be considered as an upper bound for the true one,
and it approaches the true one as $u$ grows and/or $\tau$ decreases.
Specifically, after expanding the hypergeometric function of (\ref{Rm})
(or (\ref{Rn})) in a series \cite{be}, and carrying out the integration,
we get a power series in $-1/\beta_2^2$ whose coefficient by
power $j$ depends on the hypergeometric function $F(-j,m+1/2;2m+1;1-\tau^2)$.
One can verify numerically that if $\tau \ll 1$ and $j \ll m$, then
this function is approximately equal to
$
F\ld -\!j,m\!+\!{1}/{2};\,2m\!+\!1;1 \rd
\ld 1+\tau^2  \rd^{j}.
$
On substituting this approximation to the series and summing up the series,
we arrive at an expression similar to formula (9) of Ref.\cite{jk},
which in turn constitutes an approximation to ${\cal R}_2(2m,0)$.
The obtained approximation for the correlation function
is the better the smaller is $\tau$ and/or
the larger is $u$ (see Fig.\ref{fig}). In particular,
if $\tau=0$, it becomes exact for any $\beta_2$.
On the basis of the results of Ref.\cite{jk},
the approximate asymptotic behavior of ${\cal R}_2(2m,0)$
has the form:
\bea
{\cal R}_2(2m,0)\! \approx \!
\frac{(-1)^m}{\pi\sqrt{2\beta_2}}
\frac{1}{\sqrt[4]{1\!+\!\tau^2}}\frac{\exp(-2m/\xi)}{2m} \non
\times \cos\!\ld \! m\arccos\frac{a\!-\!1}{a\!+\!1}\! -\! \frac{\pi}{4}\! \rd,
\eea
where
\bea
\xi^{-1}=\ln\lbr (\sqrt{a}+\sqrt{1+a})\sqrt{\frac{1+\tau^2}{1-\tau^2}} \rbr, \non
\label{xiappr}
a\equiv \frac{2\beta_2(\sqrt{1+\beta_2^2+\tau^2}+\beta_2)}{1+\tau^2}.
\eea
As can be seen in Fig.\ref{fig}, the above approximate $\xi^{-1}$
is a lower bound for the numerically exact one.
The asymptotic behaviors of the inverse correlation length (\ref{xiappr}),
for small and large gaps, read:
\bea
\xi^{-1}\!\approx\! \ln\sqrt{\frac{1\!+\!\tau^2}{1\!-\!\tau^2}}
+ \frac{\sqrt{2\beta_2}}{(1\!+\!\tau^2)^{1/4}}\!
\ld \! 1\!+\!\frac{1}{6}\frac{\beta_2}{\sqrt{1\!+\!\tau^2}}\! +\! \ldots \rd \! , \non
\xi^{-1}\approx
\ln\frac{4}{\sqrt{1-\tau^2}} +\ln \beta_2 +\frac{3}{16}\frac{\tau^2}{\beta_2^2}
-\frac{35}{512}\frac{\tau^4}{\beta_2^4} +\ldots ,
\eea
respectively.

To summarize, continuing our studies of the decay rates of
one-body reduced density matrices in insulators, by means of
analytic and numerical methods, we have investigated a
tight-binding model, where not only an external potential but also
hoppings can be spatially modulated.
The general result of our investigations, presented in this report
and in Ref.\cite{jk}, reads:
the form of the large-distance decay is always a power law times
an exponential,
unless the hoppings and the external potential are simultaneously uniform.
We have determined analytically the power in the power law and the
correlation length for $D=1$ system and in several directions for
$D=2$ system.
In comparison to the uniform hopping case,
the considered here system with modulated hoppings exhibits interesting
novel features.
Firstly,
since the correlation length depends on two energy parameters
(controlling the amplitudes of spatial modulation of the external potential
and the hopping), in $D=1$ case and in some directions of
$D=2$ case the gap does not determine the correlation length.
Secondly, by adjusting the hopping-modulation amplitude suitably,
the $D=2$-system can be decoupled into  quasi-one-dimensional
zig-zag  chains, which has made possible investigating a crossover
from $D=2$-decay rates to $D=1$ ones.

T.K. is grateful to the Institute of Theoretical Physics of the University of Wroc\l aw for
kind hospitality and financial support.

{}

\end{document}